\documentclass{PoS}
\usepackage{subfigure}

\title{Vector Boson production with heavy flavor quarks with the CMS experiment at LHC }

\ShortTitle{Vector Boson production with heavy flavor quarks from CMS }

\author{\speaker{Bu\u{g}ra Bilin}\thanks{On behalf of the CMS Collaboration.}\\
        Universit\'{e} Libre de Bruxelles, Brussels, Belgium \\
        E-mail: \email{bugra.bilin@cern.ch}}

\abstract{In this paper, a brief outlook on the measurements of the total and differential cross-sections of vector bosons production associated with heavy flavor quarks performed using  CMS proton - proton collision data taken at center of mass energies of 7, 8, and 13 TeV is presented. Obtained results are compared with predictions of various state-of-the-art Monte-Carlo generators.}

\FullConference{XXVI International Workshop on Deep-Inelastic Scattering and Related Subjects (DIS2018)\\
		16-20 April 2018\\
		Kobe, Japan}

\begin{document}

\section{Introduction}

Processes involving W \& Z boson production are one of the better understood processes at hadron colliders. Of those processes, the ones involving leptonic decays of W and Z bosons are among the cleanest final states experimentally, allowing studies of Quantum Chromo-Dynamics (QCD) to be carried out.

Studies of vector boson production accompanied by heavy flavor (HF) objects plays an important role at the LHC. They provide tests of perturbative QCD predictions and validate Monte-Carlo data. The measurements are also sensitive to HF parton distribution functions (PDFs). Collinear production of b quarks via gluon splitting can also be probed by studying these processes. The studies also serve as backgrounds to other Standard Model (SM) measurements and beyond the Standard Model (BSM) searches.

Measurements which are summarized in this paper cover recent vector boson and HF production results obtained by the CMS experiment \cite{Khachatryan:CMS}. The results shown are carried out in the fiducial phase space defined by correcting for detector effects as described in detail in Refs.[2-10]. 

In Section two, results of Vector boson production with b-quark jets (V + b) are presented whereas Section three covers Vector boson production with c-quark jets (V + c) results. A summary is given in Section four. 

\section{V + b measurements}

In this section, V + b results using datasets of center of mass energies of 7 and 8 TeV are presented; namely the measurements of W + b $\bar{\textrm b}$ inclusive cross section, and Z + b (b) inclusive and differential cross sections.

 To "tag" the jets originating from b-quarks, two different b-tagging algorithms, the so-called Simple Secondary Vertex (SSV) and Combined Seondary Vertex (CSV) algorithms  are used, where the chosen working point of the algorithm gives $\sim50 \%$ b-efficiency and $\sim1\%$ of mistagging rate of light flavor jets. Fig. 1 shows the distribution of secondary vertex mass, which is an input to the b-tagging algorithms.

\begin{figure}[htb!]
\centering
\subfigure[]{\includegraphics[scale=0.32]{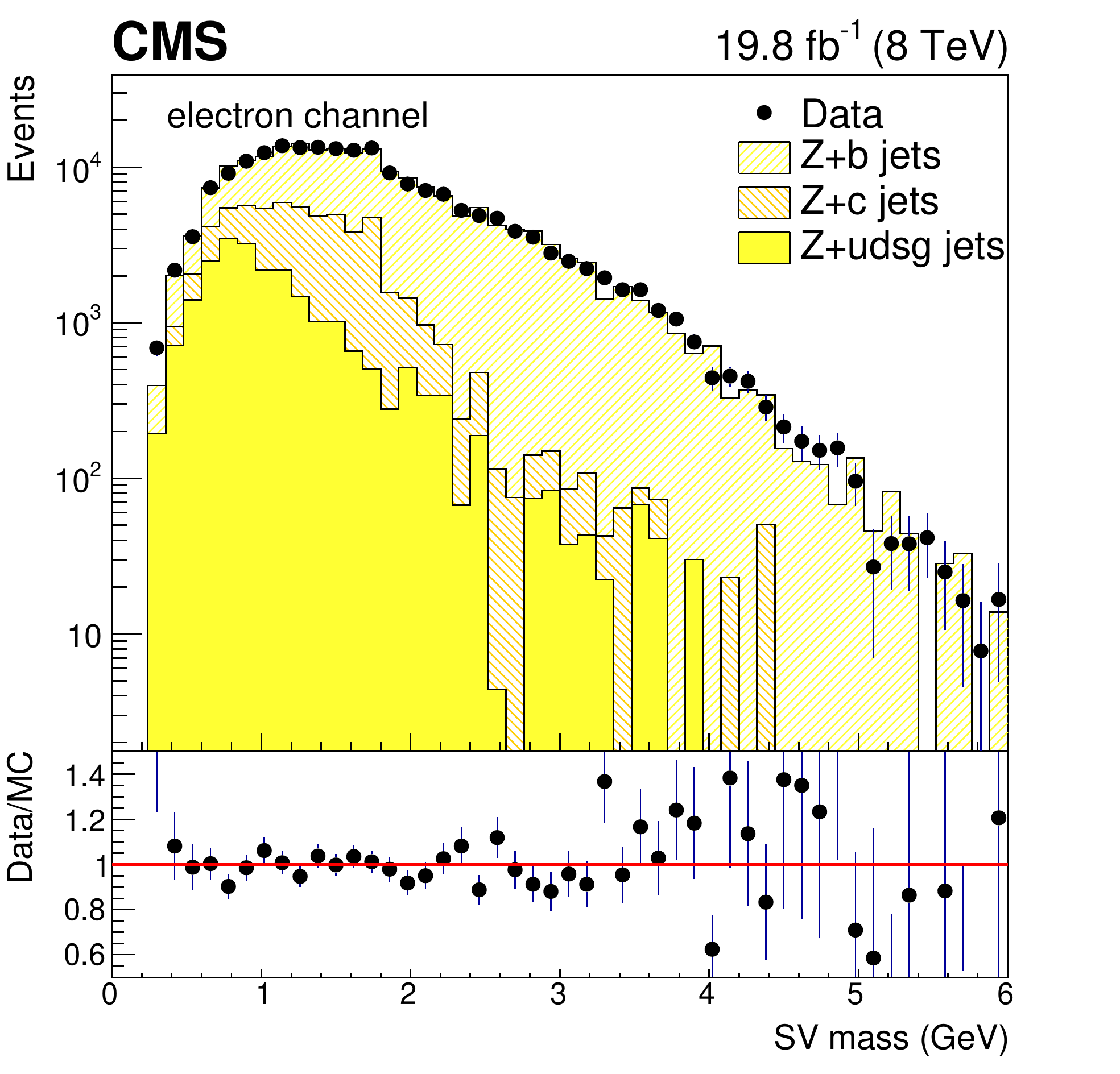}}
\hspace*{1cm}
\subfigure[]{\includegraphics[scale=0.32]{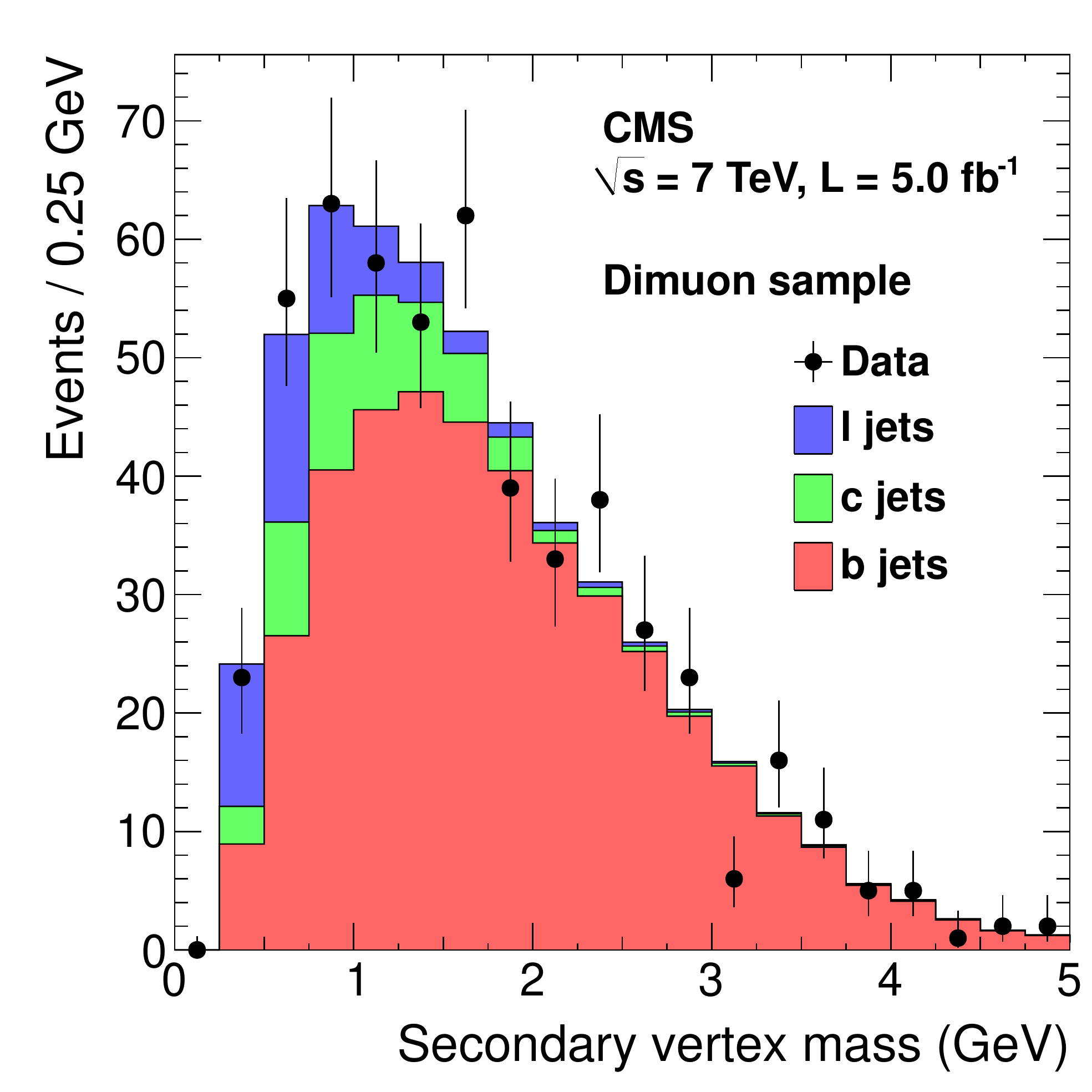}}
\caption{Distributions of the secondary vertex (SV) mass, for the leading jet after the Z + 1b selection for Z bosons decaying into di-electrons  (a) ,  and after Z + 2b selection for Z bosons decaying into di-muons  (b) \cite{Zbjet7TeV,ZBjet8TeV}. }
\label{fig:svmass}
\end{figure}

The differential cross section measurement of Z+b production is carried out with respect to various kinematic and angular variables of Z and b jets, for Z + 1b and Z + 2b categories \cite{ZBjet8TeV}. The measurements of these final states are sensitive to initial-state gluon splitting effects. The variables are also sensitive to b-quark PDFs, particularly to different PDF schemes: The five flavor scheme (5FS) which includes the b quarks in the PDFs and treats them as massless, and four flavor scheme (4FS) which does not include them and hence the b-quarks are treated massive. Fig. 2 shows the differential cross section of Z +1b production as a function of the the leading b jet transverse momentum  $p_{\mathrm T}$
 and also the ratio of the Z + 1b to Z + jets production. The shape of b jet  $p_{\mathrm T}$  is well described by the 4FS {\sc MadGraph} Leading Order (LO) prediction while the prediction underestimates the normalization by 20 \%. The shape of the leading jet $p_{\mathrm T}$ of the (Z + 1b) / (Z + jets) ratio is not well described by the 4FS based prediction. The Next-to-Leading Order (NLO) based predictions from {\sc MadGraph5}\_aMC@NLO resemble the predictions of LO {\sc MadGraph} 5FS, and  NLO calculations of {\sc powheg} have similar descriptions of the results as {\sc MadGraph5}\_aMC@NLO.  
  
 \begin{figure}[htb!]
\centering
\subfigure[]{\includegraphics[scale=0.32]{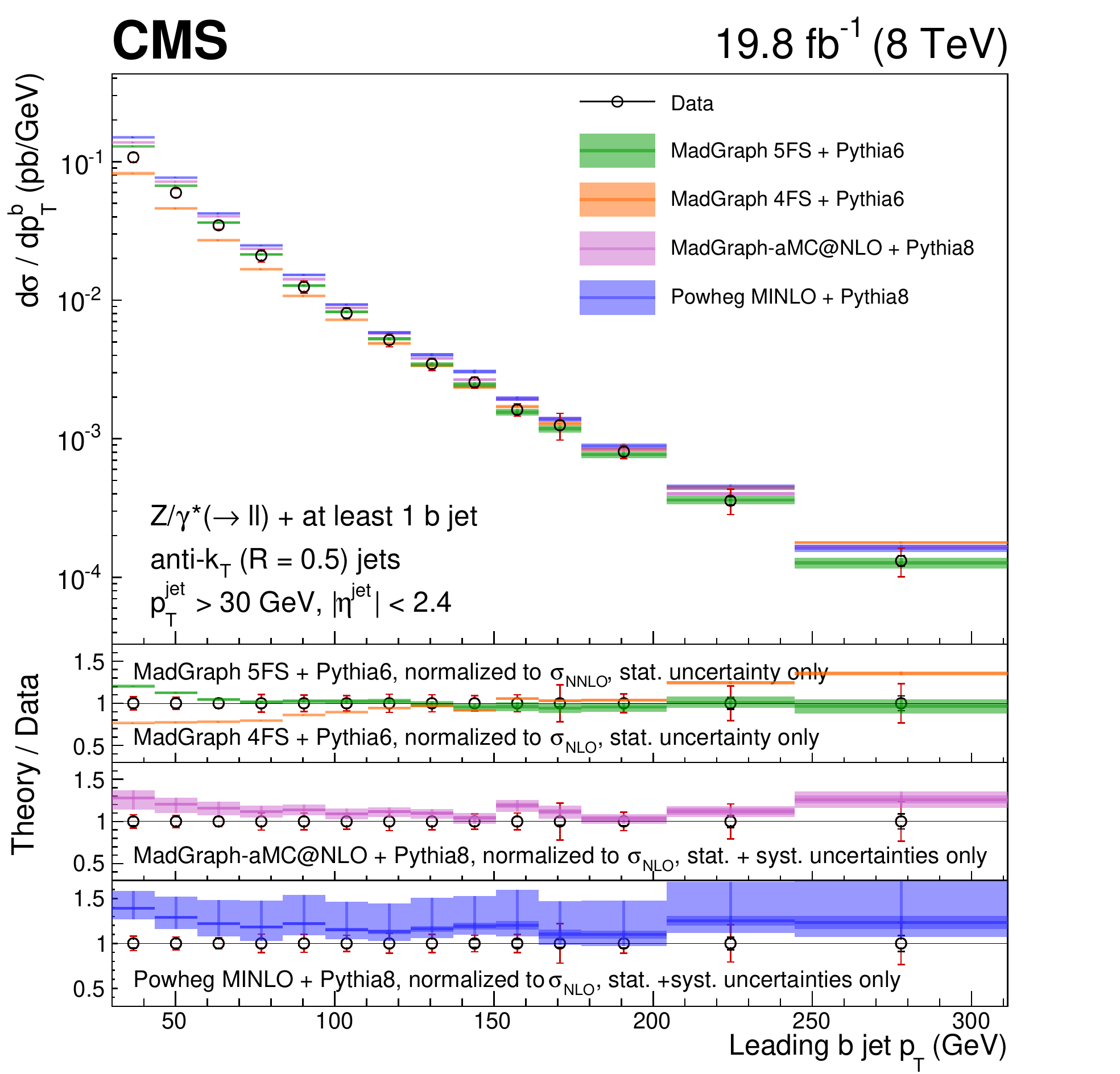}}
\hspace*{1cm}
\subfigure[]{\includegraphics[scale=0.32]{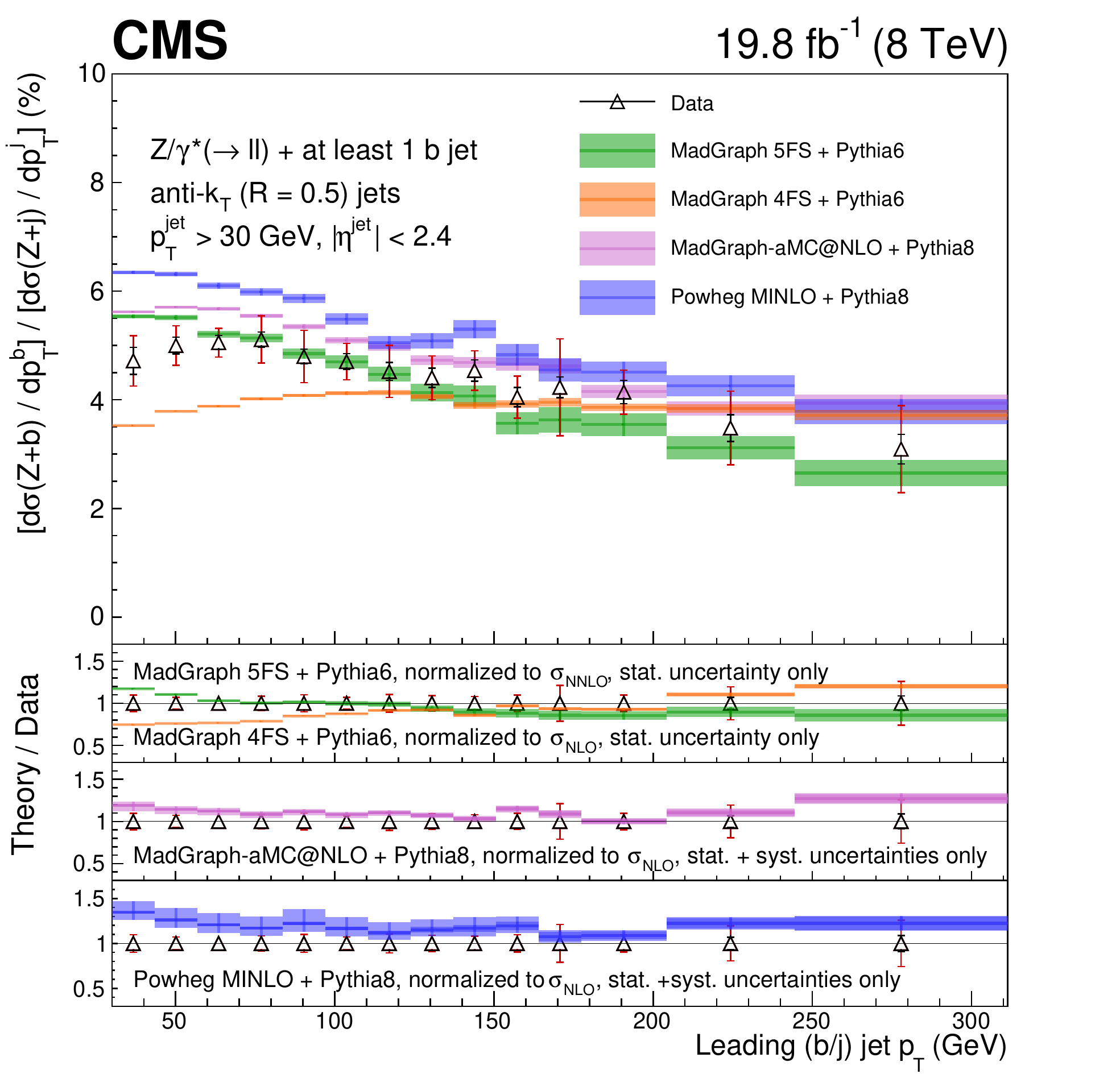}}
\caption{Differential fiducial cross section for Z +1b production as a function of the leading b jet $p_{\mathrm T}$ (a) and the cross section ratio of Z + 1b to Z + jets production as a function of leading jet $p_{\mathrm T}$  (b) \cite{ZBjet8TeV}. }
\label{fig:bptmeas}
\end{figure}

The CMS Collaboration also performed a study of Z boson production with B hadrons, where B hadrons are identified using displaced secondary vertices without any jet reconstruction \cite{Zbhad7TeV}. Hence, the measurement is not restricted to jet cones and can extend to the phase-space where the B hadrons are collinear and are sensitive to gluon splitting effects. Fig. 3 shows the differential cross section as a function of $\Delta R$ between B hadrons inclusively and also for events with Z boson transverse momentum greater than 50 GeV. A better description of the collinear region of the $\Delta R$ spectrum is obtained with the {\sc alpgen} predictions whereas 4FS and 5FS {\sc MadGraph} and a{\sc mc@nlo} predictions underestimate the data. All the generators give better description at higher $\Delta R$. For $p_{\mathrm T} (Z) > 50 $ GeV, the contribution of collinear decay is increased, hence a better description is given by the {\sc alpgen} predictions.

\begin{figure}[htb!]
\centering
\subfigure[]{\includegraphics[scale=0.32]{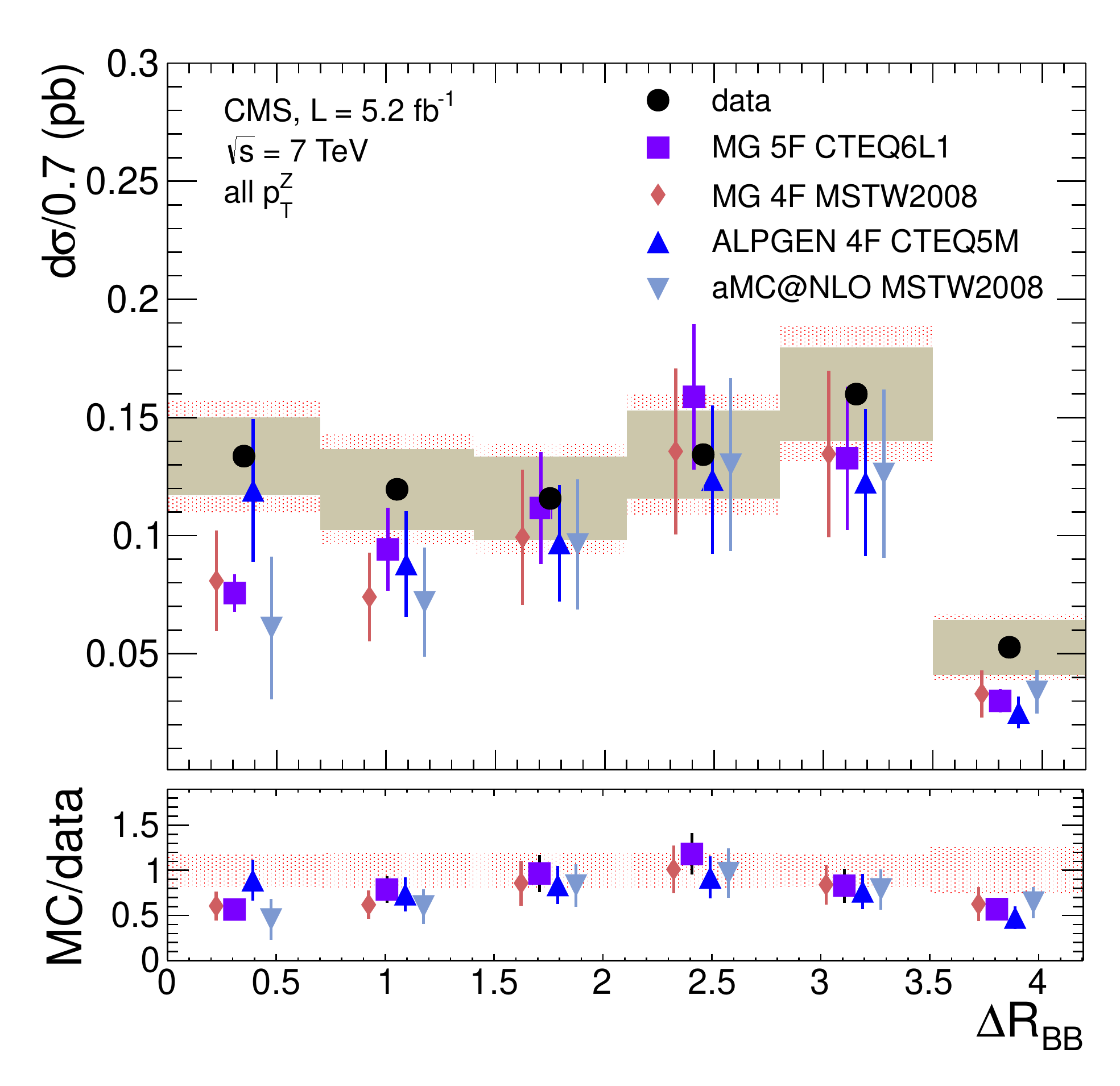}}
\hspace*{1cm}
\subfigure[]{\includegraphics[scale=0.32]{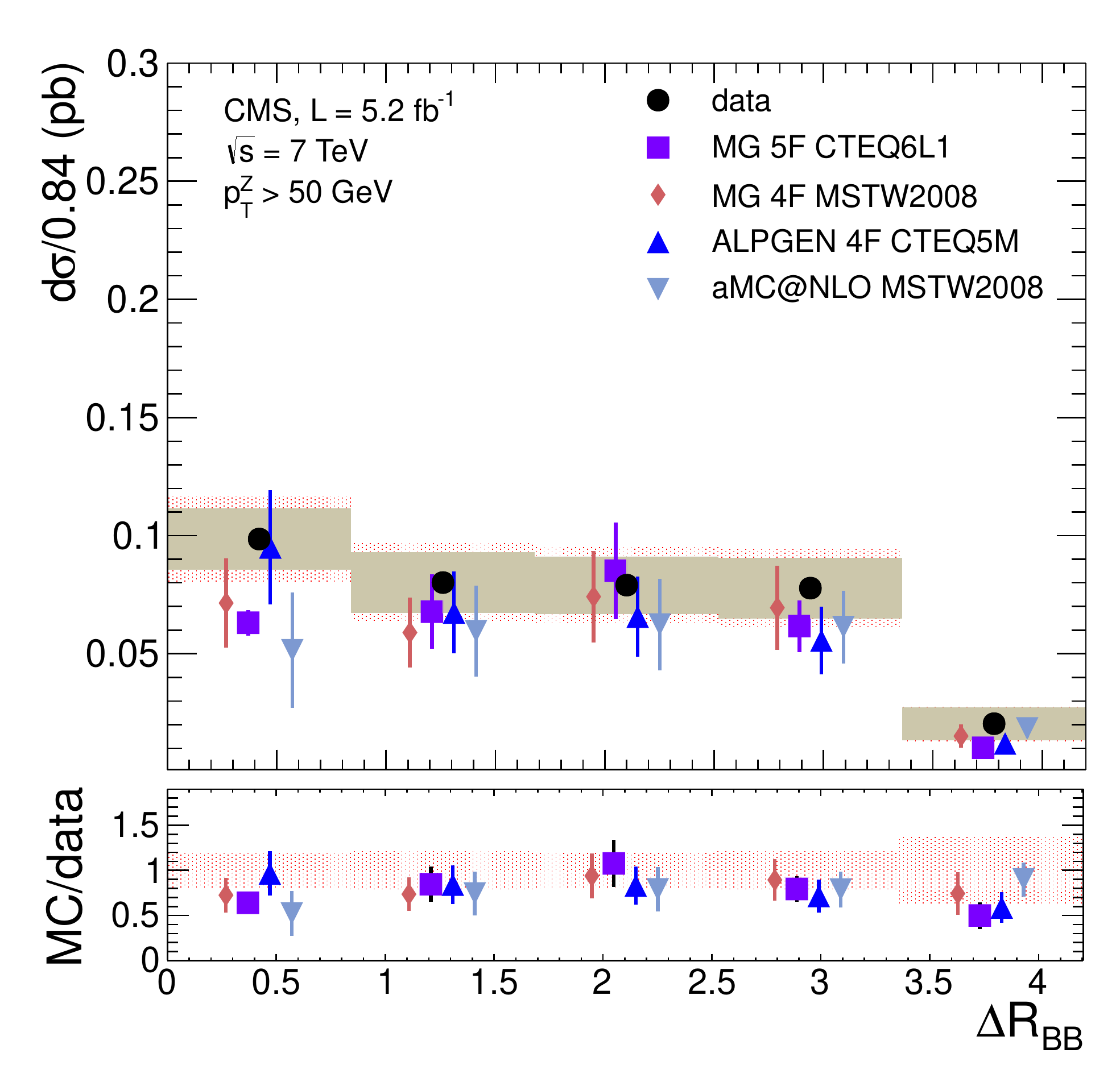}}
\caption{Differential cross sections as a function of $\Delta R _{BB}$ inclusively (a) and for $p_{\mathrm T}^{\mathrm Z} > 50 $ GeV (b) \cite{Zbhad7TeV}. }
\label{fig:bhadron}
\end{figure}

\section{V + c measurements}

V + c measurements provide constraints to strange and charm quark PDFs and they also serve as backgrounds for searches. Tagging of charm jets is carried out in 3 signatures; semileptonic decay of a hadron leading to a muon from a displaced vertex, a displaced SV with 3 tracks consistent with D$^{\pm}$ decay, and a displaced SV with two tracks consistent with D$^{0}$ decay and associated to a previous D*$^{+}$ (2010) decay.

CMS has carried out measurements of the W + c cross section as well as W$^{+}$ + $\overline{\textrm c}$ / W$^{-}$ + c ratio, and the Z + c cross section as well as Z + c / Z + b ratio. W + c measurements are done inclusively as well as differentially with respect to lepton $\eta$, whereas Z + c measurements are carried out inclusively and differentially in Z and jet $p_{\mathrm T}$. Obtained results are compared with various PDF sets. 

Fig. 4 shows the W + c production cross section at 7 and 13 TeV as a function of the lepton pseudorapidity, whereas Fig. 5 shows the cross section ratio  $\sigma$ (W$^{+}$+$\overline{\textrm c}$) / $\sigma$ (W$^{-}$+c) compared with different PDF predictions. At 7 TeV, the measurements are described well by {\sc mcfm} NLO predictions using MSTW08, CT10 and NNPDF23 PDF sets, whereas at 13 TeV,  NLO predictions obtained from {\sc mcfm} using ABMP16nlo, CT14nlo, MMHT14nlo, NNPDF3.0nlo, and NNPDF3.1nlo PDF sets are in good agreement with the data, except the prediction of ATLASepWZ16nnlo PDF. In the 13 TeV study, W+c results are further used to illustrate their impact on determining the strange quark PDFs. 

Fig. 6 shows the differential Z+c production cross section as well as the Z+c to Z+b cross section ratio as a function of $p_{\mathrm T} (Z) $. {\sc MadGraph5}\_aMC@NLO predicts higher differential cross section values than {\sc MadGraph} and {\sc mcfm} predictions with MSTW08, CT10 and NNPDF3IC PDF sets.

\begin{figure}[htb!]
\centering
\subfigure[]{\includegraphics[scale=0.31]{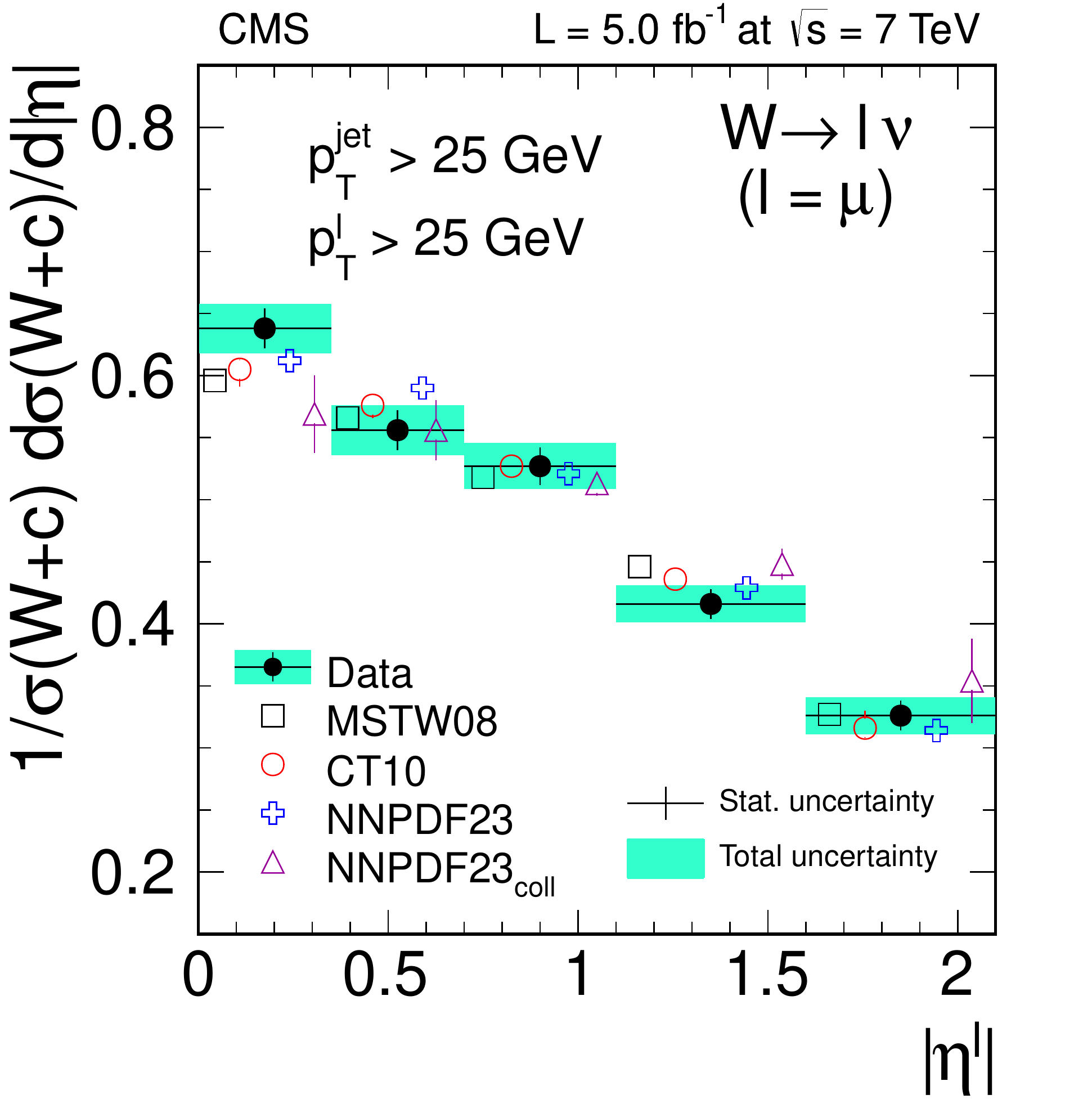}}
\hspace*{1cm}
\subfigure[]{\includegraphics[scale=0.39]{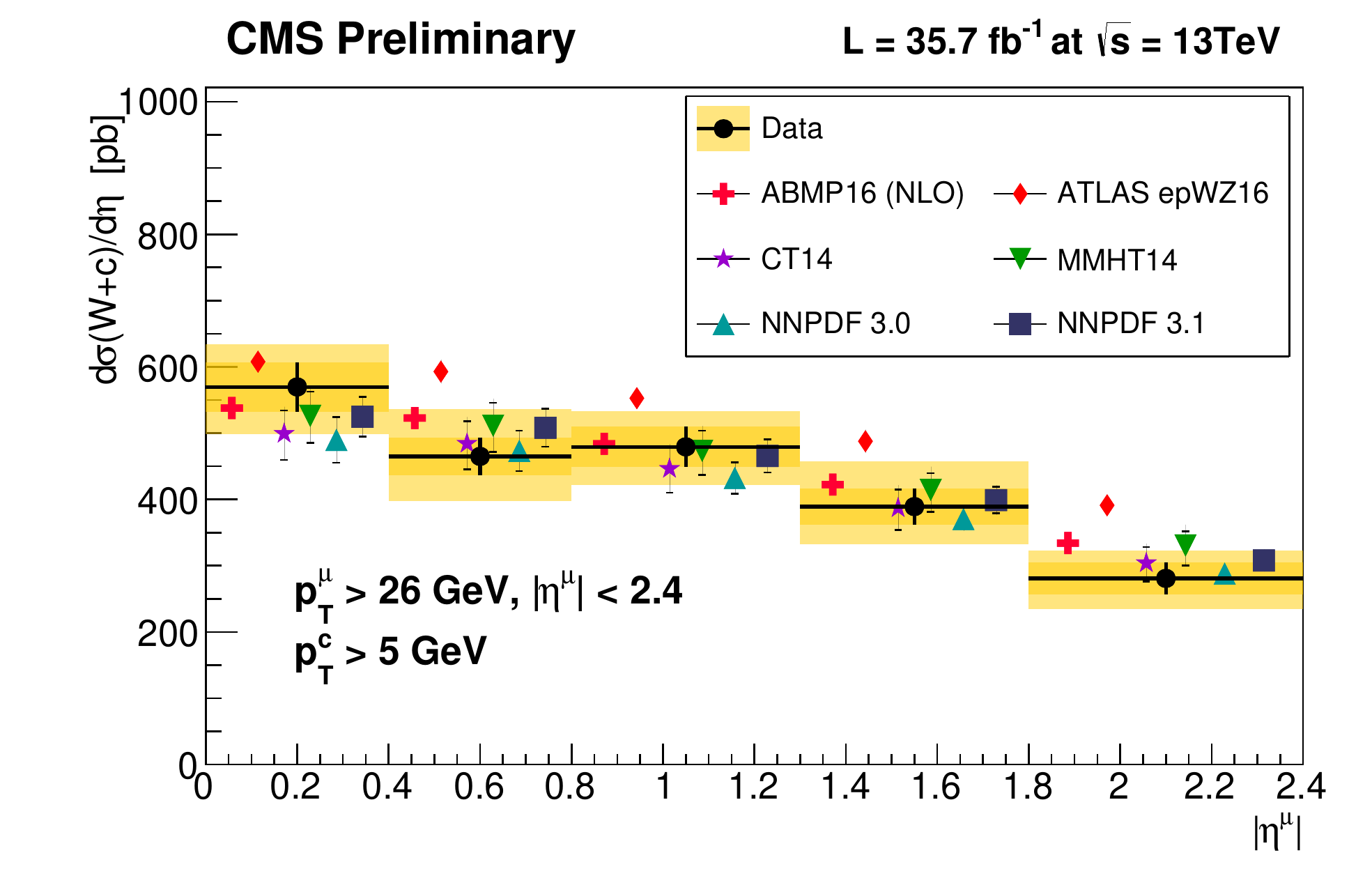}}
\caption{ W+c production cross section as a function of the pseudorapidity of the muon at 7 TeV (a) and at 13 TeV (b) \cite{Wc7TeV,PAS-SMP-17-014}. }
\label{fig:Wc}
\end{figure}

\begin{figure}[htb!]
\centering
\subfigure[]{\includegraphics[scale=0.32]{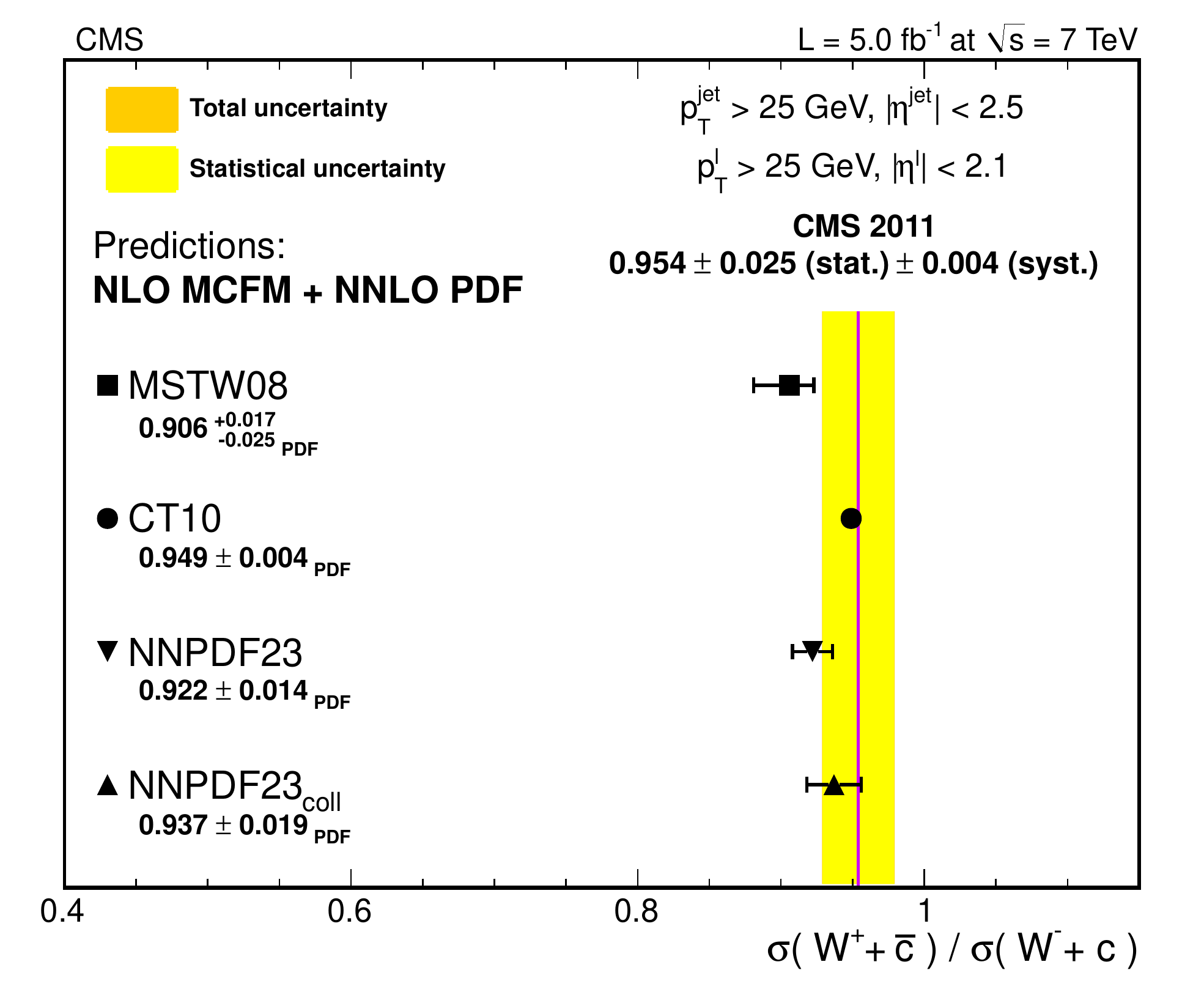}}
\hspace*{1cm}
\subfigure[]{\includegraphics[scale=0.32]{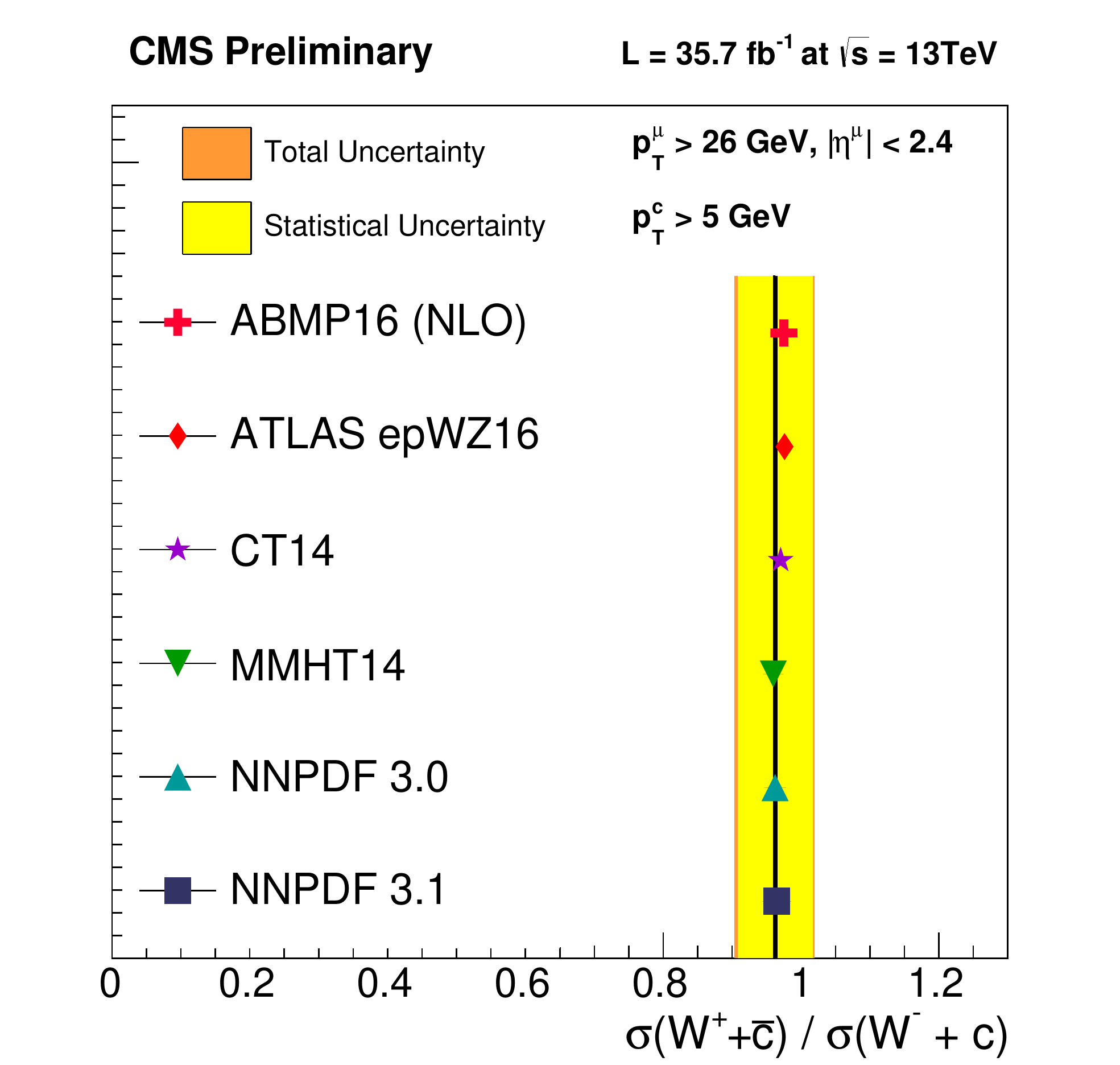}}
\caption{Cross section ratio for $\sigma$ (W$^{+}$+$\overline{\textrm c}$)/ $\sigma$ (W$^{-}$+c) at 7 TeV (a) and 13 TeV (b) compared with different PDF predictions \cite{Wc7TeV,PAS-SMP-17-014}. }
\label{fig:Wc}
\end{figure}

\begin{figure}[htb!]
\centering
\subfigure[]{\includegraphics[scale=0.32]{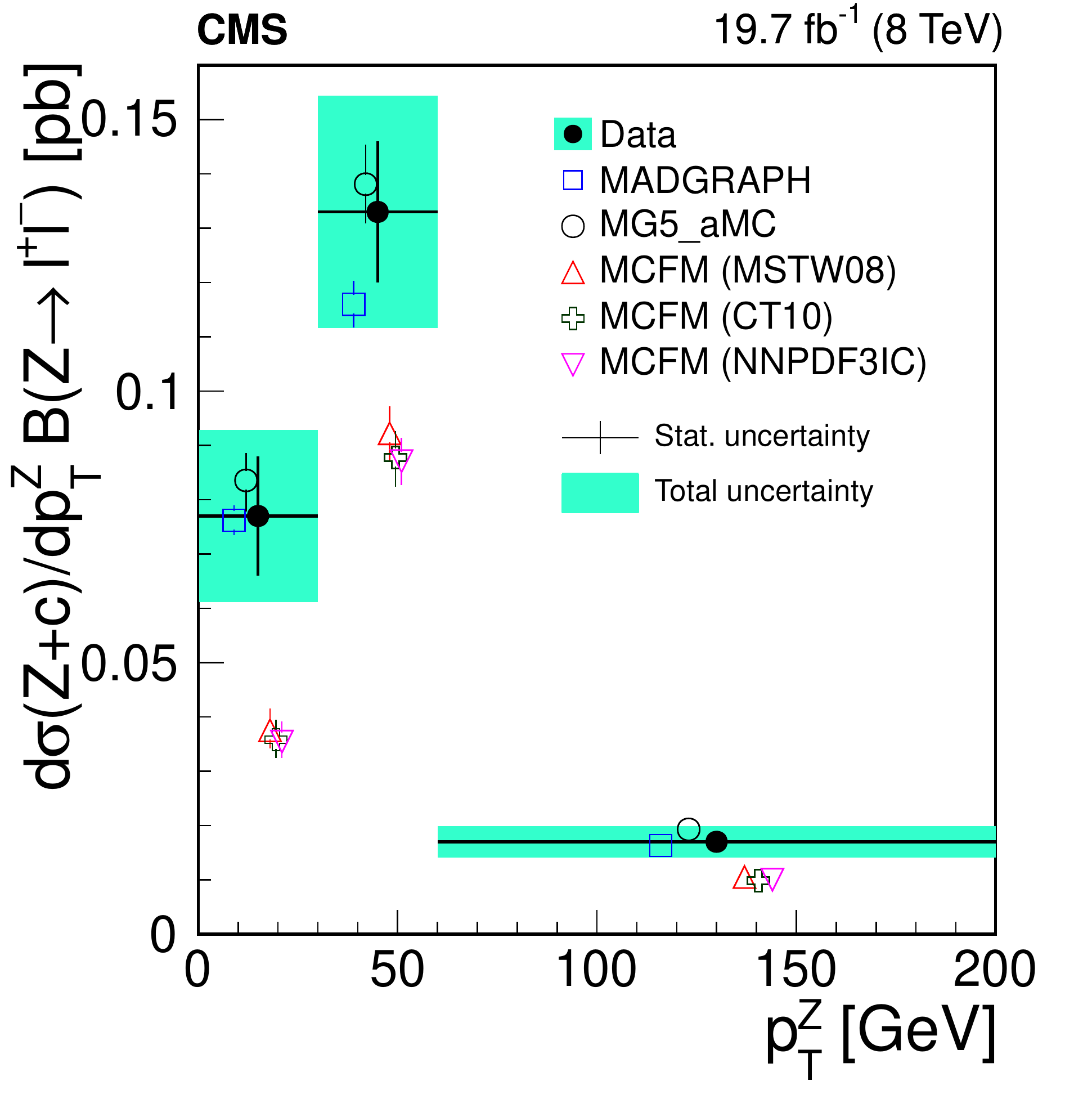}}
\hspace*{1cm}
\subfigure[]{\includegraphics[scale=0.32]{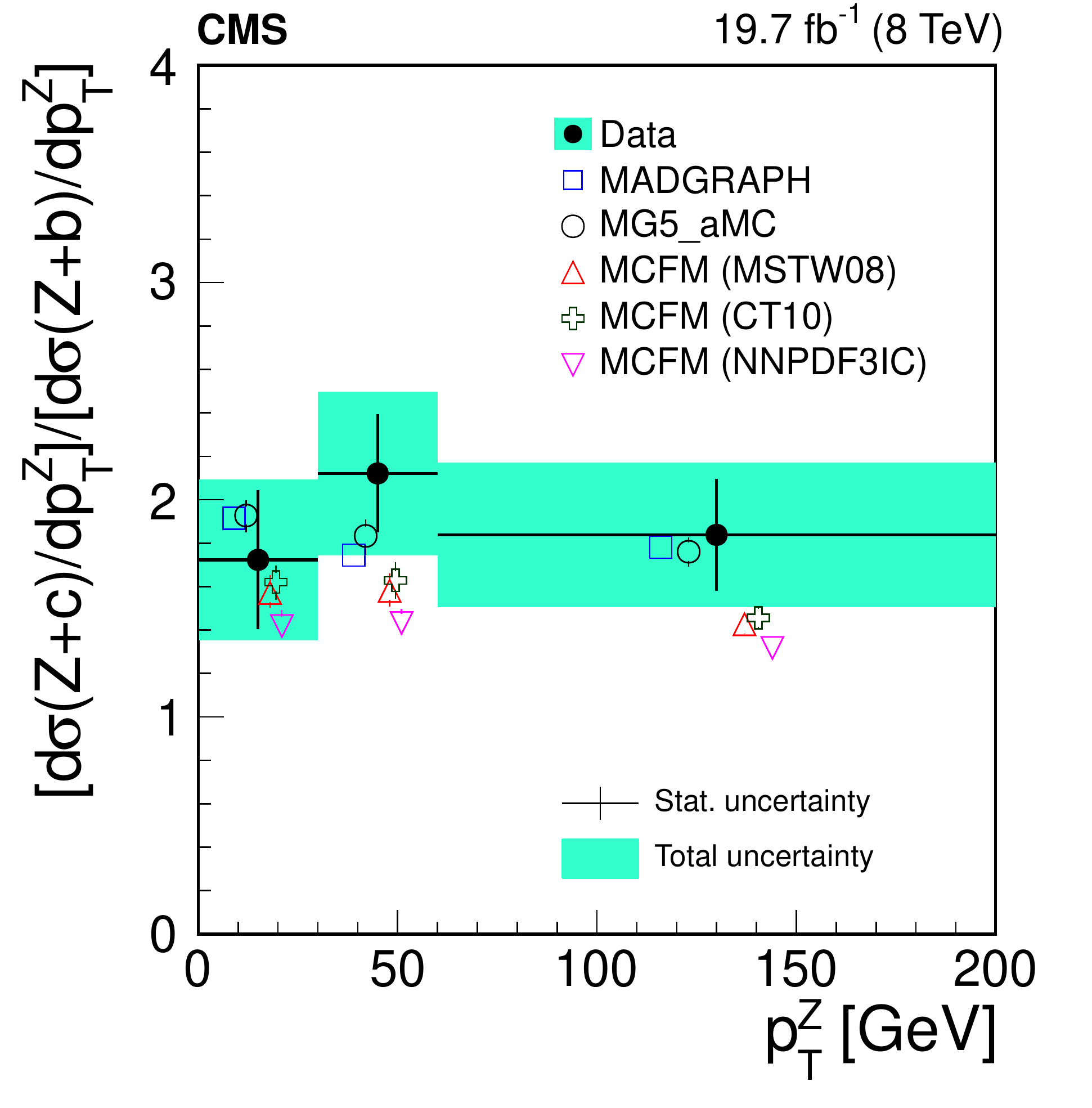}}
 \caption{Differential Z+c cross section (a) and (Z+c)/(Z+b) cross section ratio (b) measured with respect to the transverse momentum of the Z boson \cite{ZCjet8TeV}. }
\label{fig:Zc}
\end{figure}

\section{Conclusion}

The measurements presented in this note provide a detailed description of vector boson production in association with heavy flavor. They provide stringent tests for the validity of perturbative and non-perturbative QCD predictions. They also add confidence in existing Monte-Carlo models for
describing standard model data and for determining background in searches beyond the standard model.

\end{document}